\documentclass[aps,pre,preprint,groupedaddress,
,superscriptaddress,showpacs,a4paper,floatfix]{revtex4-1}

\usepackage{amsmath,amssymb,graphicx,graphics,bm}
\usepackage{hyperref}
\usepackage{graphicx,graphicx,color,ulem}
\usepackage{amsmath,amssymb}
\usepackage{epstopdf}

\definecolor{pdcolor}{rgb}{1,0.5,0}

\definecolor{pdblue}{rgb}{0,0,1}

\definecolor{rkgreen}{rgb}{0,1,0}

\definecolor{mgother}{rgb}{0,0.5,1}

\begin{document}

\title{Stochastic hydrodynamic velocity field and the representation
  of Langevin equations}

\author{Massimiliano Giona}
\email[corresponding author:]{massimiliano.giona@uniroma1.it}
\affiliation{Dipartimento di Ingegneria Chimica, Materiali, Ambiente La Sapienza Universit\`a di Roma\\ Via Eudossiana 18, 00184 Roma, Italy}

\author{Davide Cocco}
\email{davide.cocco@uniroma1.it}
\affiliation{SBAI, La Sapienza Universit\`{a} di Roma, Via Antonio Scarpa, 14,
	00161 Roma, Italy}

\author{Giuseppe Procopio}
\email{giuseppe.procopio@uniroma1.it}
\affiliation{Dipartimento di Ingegneria Chimica, Materiali, Ambiente La Sapienza Universit\`a di Roma\\ Via Eudossiana 18, 00184 Roma, Italy}

\author{Andrea Cairoli}
\email{ac2557@cam.ac.uk}
\affiliation{Centre for Mathematical Sciences, The University of Cambridge, Wilberforce Road, Cambridge CB3 0WA, United Kingdom}

\author{Rainer Klages}
\email{r.klages@qmul.ac.uk}
\affiliation{Queen Mary University of London, School of Mathematical Sciences, Mile End Road, London E1 4NS, United Kingdom}
\affiliation{London Mathematical Laboratory, 8 Margravine Gardens, London W6 8RH, United Kingdom}

\date{\today }

\begin{abstract}
The fluctuation-dissipation theorem, in the Kubo original formulation,
is based on the decomposition of the thermal agitation forces into a
dissipative contribution and a stochastically fluctuating
term. {This decomposition can be avoided by} introducing a
stochastic velocity field, {with correlation properties deriving 
from linear response theory. Here, we adopt} this field as the
comprehensive hydrodynamic/fluctuational {driver} 
of the kinematic equations of motion. 
{With this description, we show that} 
the Langevin equations for a
Brownian particle interacting with a solvent fluid become particularly
simple 
and can be applied even in those cases in which the
classical approach, based on the concept of a stochastic thermal
force, displays intrinsic difficulties 
{e.g., in the presence of the Basset force.
We show that} a convenient way for describing
hydrodynamic/thermal fluctuations is by expressing them in the form of
Extended Poisson-Kac Processes possessing prescribed correlation
properties and a continuous velocity density function. 
{We further highlight} the importance
of higher-order correlation functions in the description of the
stochastic hydrodynamic velocity field 
with special reference to short-time properties of Brownian motion. 
{We conclude by outlining some practical
implications in connection with the statistical description of particle motion in
confined geometries.}
\end{abstract}

\maketitle

\section{Introduction}
\label{sec1}

One of the major results of statistical physics for systems at thermal
equilibrium is the fluctuation-dissipation theorem, formulated by
Einstein in 1905 to 1906 in the analysis of Brownian motion
\cite{einstein}, generalized by Callen and Welton using a quantum
formalism \cite{callen} and extended by Kubo to generalized Langevin
processes \cite{kubo}. In its essence, the fluctuation-dissipation
theorem is based on the decomposition of the thermal fluctuations
into two main contributions: a dissipative force proportional to
velocity that, in the Einstein-Langevin theory, coincides with the
Stokesian friction \cite{einstein,langevin}, and a stochastic force
accounting for the thermal fluctuations.  The same setting
characterizes the Kubo approach that involves a generalized Langevin
equation of the form
\begin{equation}
\dot{v}(t) = - \int_0^t h(t-\tau) \, v(\tau) \, d \tau + \frac{R(t)}{m}
\label{eq1_1}
\end{equation}
for a particle of mass $m$ and velocity $v(t)$ in a still liquid at
thermal equilibrium, where $\dot{v}(t)=dv(t)/dt$. In
eq.~(\ref{eq1_1}), $h(t)$ is the dissipative response kernel and
$R(t)$ the stochastic fluctuating contribution.  Starting from this
formulation, Kubo generalized Einstein's result, connecting the
velocity autocorrelation function $\langle v(t) v(0) \rangle$ and the
autocorrelation function of the stochastic forcing term $\langle R(t)
R(0) \rangle$, $t \geq 0$, to the dissipative response kernel $h(t)$,
namely 
\begin{equation}
\frac{1}{\mathrm{i} \omega + h[\omega]} = \frac{1}{k_B \, T}
\int_0^\infty \langle v(t) v(0) \rangle \, e^{-\mathrm{i} \omega t} \, d t \;,
\label{eq1_2}
\end{equation}
$\mathrm{i}=\sqrt{-1}$, and
\begin{equation}
\langle R(t) R(0) \rangle =   m \, k_B \, T \, h(t)\;,
\label{eq1_3}
\end{equation}
where $T$ is the temperature, $k_B$ the Boltzmann constant and
$h[\omega]$ the Fourier-Laplace transform of $h(t)$, i.e., the Laplace
transform of $h(t)$ when the Laplace variable $s$ is set equal to
$\mathrm{i} \omega$.  {Equation~(\ref{eq1_3}) is valid if $h(t)$ is not
impulsive.} 
Equation~(\ref{eq1_2}) is customarily referred to as the first
Fluctuation-Dissipation (FD) theorem, while eq.~(\ref{eq1_3}) holds
for the second FD theorem \cite{kubo1}.

As observed by Kubo et al.\ \cite{kubo1}: ``... the random force
appearing in the fluctuation-dissipation theorem is not simple,
because the separation of the force into frictional and random forces
is itself a complex problem of statistical physics.`` This observation
indicates that some level of arbitrariness resides in this
decomposition \cite{zwanzigb}.  A similar remark is also stressed by
Tothova and Lisy \cite{tothova1}, who correctly observe that a
``physical'' thermal force, namely $R(t)$ in eq.~(\ref{eq1_1}), should
in principle be measurable independently of the hydrodynamic
interactions, while in experiments it is just a byproduct of the
reprocessing of particle trajectory time series \cite{exp1,exp2,exp3}.

The problem is even more complex for particles moving in a fluid
whenever the fluid inertia is accounted for, e.g.,  by considering the
time-dependent Stokes regime, so that eq.~(\ref{eq1_1}) is replaced by
\cite{landau,widom}
\begin{equation}
\dot{v}(t)= - \int_0^t h(t-\tau) \, v(\tau) d \tau - \int_0^t k(t-\tau) \,
\dot{v}(\tau) \, d \tau + \frac{R(t)}{m}\:.
\label {eq1_4}
\end{equation}
This equation is valid assuming $v(0)=0$, otherwise the additional
term $-k(t) \, v(0)$ should be added, as addressed in the remainder
(Section \ref{sec4}).  In eq.~(\ref{eq1_4}), the inertial effects
appear as a memory integral defined by the kernel $k(t)$ acting on the
time derivative of the velocity.  This inertial term in an
incompressible fluid is expressed by the dynamic added mass,
corresponding to the superposition of the impulsive contribution of
the added mass $m_a$ and of the Basset force \cite{maxey}
\begin{equation}
k(t)= m_a \, \delta(t) + \beta \, \frac{1}{\sqrt{t}}\:.
\label{eq1_5}
\end{equation}
For a spherical particle of radius $R_p$ it is $\beta=6 \, \sqrt{\pi \rho
  \mu} \, R_p^2$, where $\rho$ and $\mu$ are the
density and the viscosity of the fluid, respectively, and
$m_a=m\rho/(2\rho_p)$, where $\rho_p$ is the particle density.
Note that in this case, the Kubo formulation of the FD theorem does
not strictly apply \cite{kubo1}. Nevertheless, enforcing Linear
Response Theory (LRT) it is possible to recover the velocity
autocorrelation function, and out of it, the autocorrelation function
of the stochastic forcing $R(t)$ \cite{kubo1}.  Phenomenologically,
the above mentioned inertial effects correspond to the back-action to
the particle of the correlated motion of the fluid elements in its
neighbourhood \cite{darwin}.  Viewed in this perspective, it is even
more difficult to separate conceptually their influence from that of
the stochastic forcing $R(t)$.

The importance of the FD theorem is essentially two-fold: (i) to
provide a unified framework for fluctuations and dissipation, at least
for systems at thermal equilibrium. This has been originally expressed
by the Stokes-Einstein relation, connecting the main physical quantity
accounting for the intensity of thermal fluctuations, namely the
particle diffusivity $D$, to the strength of the dissipative action,
given by the Stokesian friction factor $\eta$, and to the
thermodynamic state of the system (characterised by the temperature
$T$), $D=k_B T/\eta$; (ii) to make possible direct stochastic
simulations of particle trajectories, via the concept of
Wiener-Langevin equations (with the caveat of breaking Galilei
invariance which, however, can be cured by suitably adapted methods
\cite{CKB18}). Thanks to the introduction of Wiener processes,
compactly expressing the long-term properties of Brownian fluctuations
\cite{wiener1,wiener2}, and to the seminal work by Ito on stochastic
analysis and integration \cite{ito}, this powerful numerical tool was
made available, and soon became a cornerstone in the investigation of
large-scale molecular systems starting from the analysis of individual
trajectories.  The possibility of making stochastic simulations of
particle motion paved the way to the last 70 years of research and
elaborations, not only in classical statistical physics, but also in
quantum physics (path-integrals on Wiener trajectories
\cite{feynman}), condensed-matter physics and quantum field theory
\cite{zinn}. Accordingly, new fields of physics have been opened, for
instance, the fractal theory for processes and objects controlled by
fluctuational dynamics (diffusion-limited processes, random fractals)
\cite{vicsek}, where numerical simulations played a central role for
the development of the theory of disordered and fractal systems.

Back to FD theory, out of the two main conceptual and practical
implications pointed out above, the second one is probably the most
delicate, as, within this framework, a generalized Stokes-Einstein
relation is essentially a very general and elementary property of
equilibrium and dissipation (see the discussion in Section
\ref{sec3}). The goal of this article is to show that this classical
decomposition into dissipative and fluctuational forces can be
overcome, as a comprehensive description of the hydrodynamic and
thermal fluctuations is embedded in the concept of a stochastic
hydrodynamic/thermal velocity field ${\bf v}_s(t;{\bf x})$, introduced
in this article.  Consequently, the main physical problem is the
determination and the representation of the statistical properties of
the hydrodynamic fluctuations for generic fluids and flow devices.
This approach answers the doubts and reflections of Kubo et
al.\ \cite{kubo1} on the real physical meaning of dissecting the
complex interaction of a particle with a fluid continuum into
dissipative/inertial/fluctuating contributions.  In the unitary and
compact description of the stochastic velocity field ${\bf v}_s(t;{\bf
  x})$ lies the essence of the indecomposable relation between
fluctuation, dissipation and fluid medium inertia.

In this respect, the formulation that we put forward in this paper
provides a completely different point of view compared to FD theory,
in the sense that in this approach the concept of a stochastic thermal
force is redundant. This simplifies, in turn, the application to
hydrodynamic problems, involving, e.g., fluid inertial effects. The
practical implementation of this approach for the simulation of
stochastic particle motion finally implies the reduction of the
comprehensive stochastic velocity field into generic stochastic
processes. Using Poisson-Kac processes \cite{kac,bena}, Generalized
Poisson-Kac Processes \cite{giona1,giona2,giona3} or L\'evy Walks
\cite{zaburdaev} is a simple and computionally efficient way of
expressing the stochastic velocity field for this purpose. All these
processes can be unified into the class of Extended Poisson-Kac
processes as recently formulated in Ref.~\cite{rainer}.

The article is organized as follows.  Section \ref{sec2} introduces
the velocity splitting for generic stochastic dynamics described by
equations of the form of eq.~(\ref{eq1_4}) into a ``deterministic''
velocity term, accounting for the deterministic interactions, and a
stochastic velocity field for the hydrodynamical/thermal fluctuating
contributions.  In Section \ref{sec3}, we derive the velocity
autocorrelation function of the fluctuating part using LRT.  In
Section \ref{sec4}, we describe the simple problem of a Brownian
particle influenced by inertial fluid effects.  Section \ref{sec5}
compares and contrasts the present formulation of the hydrodynamic
Langevin equation based on the definition of a stochastic hydrodynamic
velocity field with the classical approach based on eq.~(\ref{eq1_4}).
We also discuss the experimental determination of this field and how
this approach leads to a different and alternative interpretation of
FD theory. The comprehensive description of hydrodynamic/thermal
interactions within a unique stochastic velocity field shifts the
modeling focus from the classical Wiener-based description of thermal
forces to the use of stochastic processes possessing either
exponentially or power-law decaying correlation functions of time,
corresponding, in their more general setting, to the class of Extended
Poisson-Kac processes analyzed in Ref.~\cite{rainer}.  In Section
\ref{sec6}, we explore this aspect by explictly addressing how these
extended processes can be equipped, in a simple way, with a generic
velocity density function, still keeping unchanged their correlation
properties.
Furthermore, we also show how the present formulation hinges on a more
detailed description of the statistical properties of thermal velocity
fluctuations, beyond the analysis of second-order correlation
functions. Experimental validation of the theoretical predictions may
be obtained from the analysis of Brownian motion at short time scales,
using the techniques addressed in
Refs.~\cite{exp1,exp2,exp3,raizenrev1,raizenrev2}.  {We discuss the
  application of this approach to microfluidic problems in
  Appendix~\ref{sec7}. }  Apart from the practical relevance to
microfluidics, this analysis opens interesting theoretical
perspectives connected with the extension of LRT to the nonlinear
case.

\section{Deterministic-stochastic velocity splitting}
\label{sec2}

Consider eq.~(\ref{eq1_4}) in its more general setting,
\begin{equation}
\dot{\bf v} = L_d[{\bf v};{\bf x}] + L_i[\dot{\bf v};{\bf x}] + {\bf a}({\bf x})
+ \frac{{\bf R}(t)}{m}\;,
\label{eq2_1}
\end{equation}
where $L_d$ and $L_i$ are linear 
{functionals of the particle velocity,
$\{{\bf v}(\tau)\}_{\tau=0}^t$, acceleration,
$\{\dot{\bf v}(\tau)\}_{\tau=0}^t$, and position, 
$\{{\bf x}(\tau)\}_{\tau=0}^t$. 
These operators are associated
with dissipative and inertial effects.
The dependence on the position history
is relevant for studying particle motion in confined
geometries, wherein both frictional and inertial effects depend on the
particle position and are described by tensor-valued quantities
\cite{brenner,procopio}.} 
In eq.~(\ref{eq2_1}), ${\bf a}({\bf x})$ is
the acceleration deriving either from external potentials or from
externally-driven hydrodynamic flows.  Equation (\ref{eq2_1}) is
completed by the kinematic equation
$\dot{\bf x}={\bf v}$. 

{We postulate the decomposition of the particle velocity into 
a term accounting for all deterministic perturbations, ${\bf v}_d$, and 
a term accounting for all stochastic perturbations, ${\bf v}_s$.  
Accordingly, we adopt the notation with suffix ``$d$'' denoting deterministic terms and 
``$s$'' stochastic ones.
We remark that this velocity splitting technique has been 
proposed in Ref.~\cite{venditti} to study the Langevin equation of a particle moving in a tilted potential.}
We set therefore 
\begin{equation}
{\bf v}= {\bf v}_d + {\bf v}_s\;.
\label{eq2_3}
\end{equation}
Substituting this decomposition 
into eq.~(\ref{eq2_1}) and 
invoking the linearity of the operators, 
{we obtain the following}
two dynamical evolution equations,
\begin{subequations} 
        \begin{align} 
\dot{\bf v}_d &  = L_d[{\bf v}_d;{\bf x}] + L_i[\dot{\bf v}_d;{\bf x}] + {\bf a}({\bf x}) \label{eq2_4a}\\
\dot{\bf v}_s  &= L_d[{\bf v}_s;{\bf x}] + L_i[\dot{\bf v}_s;{\bf x}]  \label{eq2_4b}
+ \frac{{\bf R}(t)}{m}
        \end{align}
\end{subequations}
The two velocity contributions are coupled
via the kinematic equation, 
\begin{equation}
\dot{\bf x}(t)= {\bf v}_d(t;{\bf x}(t))+ {\bf v}_s(t;{\bf x}(t))\:.
\label{eq2_5}
\end{equation}
{We note that ${\bf v}_d(t;{\bf x})$ should be regarded as} a ``weakly perturbed''
stochastic process. Albeit its evolution equation is deterministic,
it is coupled to the stochastic term ${\bf v}_s(t,{\bf x})$ via 
eq.~(\ref{eq2_5}).  

Equations~(\ref{eq2_4a}),~(\ref{eq2_4b}), 
~(\ref{eq2_5}) deserve further discussion and
explanation, as they represent the core of this, otherwise elementary,
transformation of variables.  In the transformed system of variables,
the phase-space independent coordinates of the particle are its
position vector ${\bf x}(t)$ and its ``deterministic'' velocity
component ${\bf v}_d(t;{\bf x}(t))$. These are stochastic processes
dependent on time $t$. 
{The interpretation of the term ${\bf v}_s(t;{\bf x})$ 
is altogether different.  
In essence, it is a stochastic field of fluctuations providing 
the overall 
description of the hydrodynamic/thermal fluctuations
exerted by the fluid/molecular environment on the particle. 
In practice, it is a stochastic process, governed by
eqs.~(\ref{eq2_4a}),(\ref{eq2_4b}), whose statistical properties 
are modulated by the position ${\bf x}$. } 

{The statistical description of this field} involves the characterization of its properties 
conditional to a fixed value of the position. 
This approach is identical to the determination of the position-dependent friction tensor in microfluidic channels for fixed ${\bf x}$, 
and the equations of motion follow by superimposing to the so-determined friction forces the influence of external perturbations, 
while assuming that the latter do not modify the former \cite{brenner}. 
The validity of this method relies on the ansatz that the statistical structure of 
  ${\bf v}_s(t;{\bf x})$ does not depend on 
  the externally forcing deterministic perturbations, 
  gathered in the term ${\bf a}({\bf x})$.

Conversely, if both $L_d$ and $L_i$ do not depend on the particle position ${\bf x}$, 
a situation that occurs in free space or in microfluidic channels far away from solid boundaries,
${\bf v}_s(t)$ is {independent on ${\bf v}_d$}. 
In consequence, it is simply prescribed as a stochastic process in time. 


\section{Linear Response Theory}
\label{sec3}

{In this section, we show how to derive}
the statistical properties of ${\bf v}_s(t;{\bf x})$. 
We consider the statistics of this field conditional to a given value of the position
vector ${\bf x}$. 
Henceforth, 
{to simplify the notation, we omit any}
explicit dependence on ${\bf x}$.  
From LRT
\cite{kubo,kubo1,zwanzigb,politi,reichl}, the correlation function of
${\bf v}_s$ can be derived by considering the linear response of the
dynamics to an initial condition ${\bf v}_s^0$, averaging it with
respect to the equilibrium measure of ${\bf v}_s^0$ by assuming
independence between ${\bf R}(t)$ and ${\bf v}_s^0$.  As observed by
Tothova et al. \cite{tothova2,tothova3} the original formulation of
this approach is due to V.~Vladimirsky \cite{vladimirsky} in a
scarsely known paper from 1942 written in Russian. 
Equation~(\ref{eq2_4b}) can thus be rewritten in
operatorial form as
\begin{equation}
(I- L_i) \dot{\bf v}_s(t) = L_d {\bf v}_s(t) + \frac{{\bf R}(t)}{m}\;,
\label{eq3_1}
\end{equation}
where $I$ is the identity operator,
equipped with the initial condition ${\bf v}_s(t=0)={\bf v}_s^0$. The solution
of eq.~(\ref{eq3_1}) is
\begin{equation}
{\bf v}_s(t) = e^{(I-L_i)^{-1} L_d t} {\bf v}_s^0+ \frac{1}{m} e^{(I-L_i)^{-1} L_d t} * {\bf R}(t)\;,
\label{eq3_2}
\end{equation}
where ``$*$'' denotes the convolution operation.  Indicating with
$v_{s,h}(t)$ the $h$-entry of ${\bf v}_s(t)$, it follows that
$v_{s,h}(t) v_{s,k}(0)$ can be expressed as $v_{s,h}(t) v_{s,k}(0) =
\sum_j \left (e^{(I-L_i)^{-1} L_d t} \right )_{h,j} v_{s,j}(0)
v_{s,k}(0) + \sum_j \left (e^{(I-L_i)^{-1} L_d t} \right )_{h,j} *
R_{j}(t) v_{s,k}(0)$. If $\langle \ldots \rangle$ holds for the average
with respect to the equilibrium probability measure for the velocities
and random fluctuations, from the independence of ${\bf R}(t)$ and
${\bf v}_0$ it follows that $\langle R_{j}(t) v_{s,k}(0) \rangle=0$.
Moreover,
\begin{equation}
\langle v_{s,j}^0 v_{s,k}^0 \rangle = \frac{k_B T}{m} \, \delta_{j,k}\;.
\label{eq3_3}
\end{equation}
{
We remark that added mass effects can be relevant in experimental settings \cite{raizen1}. 
These can be accounted for in the formalism by adding the added mass term to the particle mass $m$.}  
In any case, the entries of the velocity autocorrelation
tensor can be expressed as
\begin{equation}
\langle v_{s,h}(t) v_{s,k}(0) \rangle =  \frac{k_B T}{m} \,.
 \left (e^{(I-L_i)^{-1} L_d t} \right )_{h,k}
\label{eq3_4}
\end{equation}
In tensorial form, therefore, the velocity autocorrelation
function attains the expression 
\begin{equation}
{\bf C}^{v_s}(t) = \langle {\bf v}_s(t) \otimes {\bf v}_s(0) \rangle
= \frac{k_B T}{m}  \, e^{(I-L_i)^{-1} L_d t}\;.
\label{eq3_5}
\end{equation}

We now restore all the functional dependencies on the position ${\bf x}$. 
From eq.~(\ref{eq3_5}), it follows that the  
entries ${C}^{v_s}_{h,k}(t \, | \, {\bf x})$, $h,k=1,2,3$,
of the  conditional correlation
tensor ${\bf C}^{v_s}(t \, | \, {\bf x})$
  for ${\bf v}_s(t;{\bf x})$ at a given
position ${\bf x}$ can be expressed as
\begin{equation}
{C}^{v_s}_{h,k}(t \, | \, {\bf x}) =  \frac{k_B T}{m}
\, c_h^{(k)}(t\, | {\bf x}) 
\label{eq3_6}
\end{equation}
where $c_h^{(k)}(t\, | \, {\bf x})$, for fixed $k$,
 are the entries of the vector-valued
function ${\bf c}^{(k)} (t \, | \, {\bf x})$ satisfying the initial
value problem
\begin{subequations} \label{eq3_7}
        \begin{align}
\dot{\bf c}^{(k)}(t\, | \, {\bf x})& = L_d[{\bf c}^{(k)}(t\, | \, {\bf x});{\bf x }]
+ L_i[{\bf c}^{(k)}(t\, | \, {\bf x});{\bf x}] \\
c_h^{(k)}(0\, | \, {\bf x})& = \delta_{h,k}\:.
        \end{align}
\end{subequations}
The solution of eqs.~(\ref{eq3_7}) determines the spatio-temporal
correlation properties of the stochastic velocity field ${\bf
  v}_s(t;{\bf x})$. Equations~(\ref{eq3_7}) are analogous to the
evolution equation for the correlation function deriving from the
original Vladimirsky approach that considers, instead of $c(t \, |
{\bf x})$, the integral of the correlation function $V(t \, |\, {\bf
  x} ) = \langle v^2 \rangle \int_0^t c(\tau \, | \, {\bf x}) d \tau$
(see eq.~(6) in \cite{tothova1} and the related discussion).

A final comment concerns generalized Stokes-Einstein FD
relations. Consider eq.~(\ref{eq1_4}) for a Brownian particle in a
still fluid in the scalar approximation, motivated by the isotropy of
the problem.  In this case $v_s(t)=v(t)$
and $v_d=0$.
Let $\widehat{h}(s)$, $\widehat{k}(s)$ be the Laplace transforms of
the dissipative and inertial memory kernels, respectively. These two
functions satisfy the following properties: (i)
$\widehat{h}(0)=\int_0^\infty h(t) \, dt=\eta_\infty >0$, where
$\eta_\infty$ is the effective friction factor of the model that we assume it is bounded.
(ii) $\lim_{s \rightarrow 0} s \widehat{k}(s)=0$, i.e., 
there are no dissipative contributions. 
We also assume that $k(t)$ does not
contain any impulsive contribution {(i.e., no added mass)} 
corresponding to
the condition $\lim_{\varepsilon \rightarrow 0} \int_0^\varepsilon
k(t) \, d t= 0$. 
This latter condition can be easily removed, 
because it does not alter the final result. 
From eqs.~(\ref{eq3_7}), the normalized correlation function $c(t)$
admits a Laplace transform $\widehat{c}(s)$ as a solution of the
equation
\begin{equation}
m \, s  \, \widehat{c}(s) - m = - \widehat{h}(s) \, \widehat{c}(s)
- \widehat{k}(s) \,  s \,  \widehat{c}(s)
\label{eq3a_1}
\end{equation}
and thus
\begin{equation}
\widehat{c}(s)= \frac{m}{m \, s + \widehat{h}(s)+ s \, \widehat{k}(s)}\:.
\label{eq3a_2}
\end{equation}
The diffusion coefficient $D$ is the time integral of the correlation
function $\langle v(t) \, v(0) \rangle=\langle v_s(t) \, v_s(0) \rangle$ 
and is thus equal to the value $\widehat{c}(s=0)$, namely
\begin{equation}
D= \int_0^\infty \langle v_s(t) v_s(0) \rangle \, dt =
\langle v^2 \rangle \int_0^\infty c(t) \, dt =
\langle v^2 \rangle  \, \widehat{c}(0) = \frac{\langle v^2 \rangle \, m}{\eta_\infty}
\label{eq3a_3}\:.
\end{equation}
Further assuming $\langle v^2 \rangle= k_B \, T/m$, 
we recover the generalized Stokes-Einstein relation
$D= \frac{k_B \, T}{\eta_\infty}$. 
In our theory, this follows as a consequence of bounded friction.

{\section{A simple example: Brownian motion in an inertial fluid}}
\label{sec4}

We consider a Brownian particle
in free space subjected to hydrodynamic interactions including fluid
inertia. This example is not only interesting in itself, but is also instructive 
to clarify a common source of misunderstanding \cite{widom}. 
From time-dependent Stokes hydrodynamics, the Laplace
transform of the force $\widehat{\bf F}_{f \rightarrow p}(s)$ exerted
by a Newtonian fluid on a spherical particle of radius $R_p$ ($s$ is
the Laplace variable) is given by \cite{kim}
\begin{equation}
\widehat{\bf F}_{f \rightarrow p}(s) = - 6 \pi \mu R_p \, \widehat{\bf v}(s)
- 6 \pi \sqrt{\rho \mu} R_p^2  \, \frac{1}{\sqrt{s}} \, (s \, \widehat{\bf v}(s))
- \frac{2}{3} \rho \pi R_p^3 \, (s  \, \widehat{\bf v}(s))\:.
\label{eq4_1}
\end{equation}
{Transforming this equation back to the time domain 
and neglecting the fluctuations ${\bf R}(t)$, 
we obtain the evolution equation}
\begin{equation}
m \,\dot{\bf v}(t)= - \eta \, {\bf v}(t) - \beta \frac{1}{\sqrt{t}} * \left [ 
\dot{\bf v}(t)+  {\bf v}(0) \delta(t) \right ] 
\:,
\label{eq4_2}
\end{equation}
where $\eta=6 \pi \mu R_p$, $\beta=6 \sqrt{\pi \rho \mu} R_p^2$ and
${\bf v}(0)={\bf v}(t=0)$. 
{For the sake of simplicity, we neglect in this equation the added mass term. 
  While this is of physical relevance in other contexts,
  it is not important for our current discussion.} 

The dissipative and inertial functionals, $L_d$ and $L_i$,
are defined as 
\begin{subequations} \label{eq4_3bis}
        \begin{align}
L_d[{\bf v}] &=  - \eta \, {\bf v} \\
L_i[\dot{\bf v}] & = - \beta \frac{1}{\sqrt{t}} * \left [ 
\dot{\bf v}(t)+  {\bf v}(0) \delta(t) \right ] .
        \end{align}
\end{subequations}

For the following discussion it suffices to restrict ourselves to the one-dimensional case. 
Let $c(t)= \langle v(t) v(0) \rangle/\langle
v^2 \rangle$ so that $c(0)=1$ and eq.~(\ref{eq4_2}) becomes
\begin{equation}
m \, \dot{c}(t)  = - \eta \, c(t) - \beta \int_0^t \frac{1}{\sqrt{t-\tau}} \, \frac{d c(\tau)}{d \tau} d \tau - \frac{\beta}{\sqrt{t}}\:.
\label{eq4_3}
\end{equation}
Equation~(\ref{eq4_3}) coincides with the analogous relation obtained
by Widom, eq.~(9) in \cite{widom}, 
Observe that LRT does not provide the estimate for
$\langle v^2 \rangle$, which should be derived from
kinetic/hydrodynamic arguments.

\begin{figure}
\includegraphics[width=10cm]{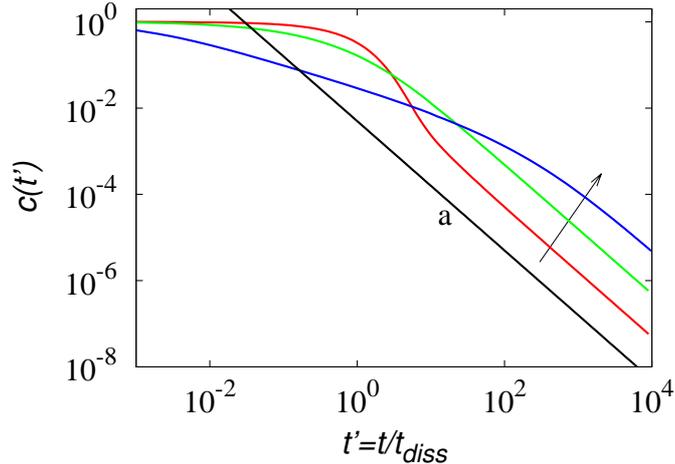}
\caption{Normalized autocorrelation function $c(t^\prime)$ vs
  $t^\prime=t/t_{\rm diss}$ for a spherical particle in a still fluid,
  the dynamics of which is defined by eqs.~(\ref{eq4_2}). 
  The arrows indicate increasing values of the nondimensional
  parameter $\gamma=10^{-1},1,10$. Line (a) represents the long-term
  scaling $c(t^\prime) \sim (t^\prime)^{-3/2}$.}
\label{Fig1}
\end{figure}

Introducing a dimensionless time $t^\prime=t/t_{\rm diss}$, rescaled
with respect to the dissipation time $t_{\rm diss}=m/\eta$,
eq.~(\ref{eq4_3}) becomes 
\begin{align}
\dot{c}(t^{\prime}) &= - c(t^{\prime}) - \gamma 
\int_0^{t^\prime} \frac{1}{\sqrt{t^\prime-\tau}} \, \frac{d c(\tau)}{d \tau} d \tau - \frac{\gamma}{\sqrt{t^\prime}} 
\qquad \text{where} \qquad 
\gamma = \left ( \frac{9}{2 \, \pi} \, \frac{\rho}{\rho_p} \right )^{1/2}
= \frac{t_{\rm diss}}{t_{\rm inert}}
\label{eq4_4}
\end{align}
expresses the ratio of the dissipation time $t_{\rm diss}$ to the
characteristic time $t_{\rm inert}$ for the occurrence of inertial
effects.  For $\rho= 10^3 \rho_p$ (such as for gas bubbles in water),
$\gamma=38$, while for $\rho_p=5 \cdot 10^3 \rho$ (heavy solid
particles in air), $\gamma=1.7 \cdot 10^{-2}$, so that the physical
range of values of $\gamma$ is $(10^{-2},10^2)$.  Figure~\ref{Fig1}
shows the behaviour of $c(t)$, obtained by solving eq.~(\ref{eq4_3}),
in non-dimensional form for several values of $\gamma$.

Apart from the well-known $t^{-3/2}$ long-term scaling induced by the
effect of the Basset force, and typical for Brownian motion in liquids
\cite{bliquid1,bliquid2,bliquid3,bliquid4,
  bliquid5,bliquid6,bliquid7,bliquid8}, Fig.~\ref{Fig1} indicates that
the influence of $\gamma$ is significant in modulating the short-time
behaviour of the velocity autocorrelation function. This observation
will be further addressed in Sec.~\ref{sec6}.

\section{Fluctuation-dissipation theory: a change of  perspective}
\label{sec5}

In the previous sections, we have developed the formalism leading to
the formulation of the hydrodynamic Langevin equations for particle
motion in a fluid medium in terms of the stochastic velocity field
${\bf v}_s(t)$. The formal simplicity of the approach, based on the
velocity splitting and on LRT, may suggest that it would represent an
alternative ``reshuffling'' of known concepts without any major
novelty, nor significant physical meaning. 
Here, we compare and contrast the proposed approach 
against the existing FD theory, showing not only its
relevance for cases of physical importance, but also how to modify the
perception of known FD relations.

The FD analysis of Langevin equations has been developed by
considering dissipative memory effects, i.e., with reference to
eq.~(\ref{eq1_1}).  The hydrodynamic approach to Brownian motion in
fluids made clear lately that fluid inertia (deriving from a
time-dependent Stokesian analysis of fluid-particle interactions)
represents an important correction at short times to the long-term
picture based on instantaneous dissipation \cite{landau,widom}. 
{Consider 
eq.~(\ref{eq1_4}) for a spherical particle of radius $R_p$ in a Newtonian fluid with viscosity
$\mu$ and density $\rho$.  
In this case, $h(t)=\eta \delta(t)$, whereas $k(t)$ is given by
eq.~(\ref{eq1_5})}. 
{This problem has been investigated extensively in experiments, 
which have shown the importance of the fluid inertia
in determining the short-time dynamics of these Brownian particles
\cite{exp1,exp2,exp3,raizenrev1,raizenrev2}.}
The presence of memory
terms, depending on the history of the particle acceleration, makes
the traditional approach based on Kubo FD theory \cite{kubo}
technically impossible, as correctly observed in \cite{kubo1}. 
Nevertheless, several authors \cite{low,bedeaux} have analyzed the
representation of the thermal force $R(t)$ in the presence of
fluid-inertial effects. 
Specifically, Bedeau and Mazur \cite{bedeaux} derived for the Fourier
transform $\widehat{R}(\omega)$ of the thermal force 
the following result:
\begin{equation}
\langle \widehat{R}(\omega) \, \widehat{R}(\omega^\prime)= 2 \, k_B \, T Re[\widehat{\zeta}(\omega)]
\, \delta(\omega-\omega^\prime)  = \widehat{\zeta}_R(\omega) \, \delta(\omega-\omega^\prime)  \:,
\label{eqbe1}
\end{equation}
where
\begin{equation}
\widehat{\zeta}(\omega) = 6 \, \pi \, \mu  \, R_p \left [1+(1- \mathrm{i} ) R_p (\omega \, \rho/(2 \, \mu))^{1/2} - (\mathrm{i} \, \omega \, \rho \, R_p^2/9 \, \mu) \right ]\:.
\label{eqbe2}
\end{equation}
{Equations~(\ref{eqbe1}),(\ref{eqbe2}) seem 
to suggest that 
Kubo fluctuation-dissipation theory can be applied also in this setting. 
However, we show below that this is ultimately problematic.} 

From eqs.~(\ref{eqbe1}),(\ref{eqbe2}) it follows that the correlation
function $\langle R(t) \, R(0) \rangle$ is the inverse Fourier
transform $\zeta_R(t)$ of $\widehat{\zeta}_R(\omega)$ entering
eq.~(\ref{eqbe1}).  
{$\widehat{\zeta}(\omega)$ can be further simplified by 
renormalizing the term $\propto \omega$ into the added mass. 
The equation then reduces to
\begin{equation}
\widehat{\zeta}(\omega) = 6 \, \pi \, \mu  \, R_p \left [1+(1- \mathrm{i} ) R_p (\omega \, \rho/(2 \, \mu)^{1/2} \right ]\:.
\label{eqbe2a}
\end{equation}
}
This relation can be used to derive scaling results for the thermal
forces, such as
\begin{equation}
\langle R(t) \, R(0) \rangle \sim t^{-3/2}
\label{eqbe3}
\end{equation}
near $t=0$, as discussed in \cite{exp2}, which are consistent with
experimental results for the power spectral density of the thermal
noise \cite{exp1,exp2,exp3}.  However, it is important to observe that
the power spectral density of $R(t)$ is a result of the
post-processing of the experimental measurements of position and
velocity of a Brownian particle. 
For this post-processing, a hydrodynamic model must be assumed. 
Therefore, $\langle R(t) \, R(0)
\rangle$, or its Fourier transform, is not a directly measurable
quantity.

{In any case, due to the power-law singularity of eq.~(\ref{eqbe3}) near $t=0$, 
we can show that there is no stochastic process possessing
eqs.~(\ref{eqbe1})-(\ref{eqbe2}) as the Fourier transform of its
autocorrelation function (see Appendix~B).  
As such, these equations should be viewed as purely formal. 
In addition, we can also show that the 
fluid-inertial interactions expressed by the Basset term, leading to eq.~(\ref{eqbe3}), renders
the representation of the thermal force unphysical and difficult to
handle in practical Langevin simulations (Appendix~B). 
This is not the case in the presence of only purely dissipative hydrodynamic effects (Appendix~A).  
%
These observations altogether supports our view that standard FD theory is only an approximate description for Brownian dynamics in more general settings.}

The velocity splitting approach in which the characterization
of the thermal fluctuations is described by means of the stochastic
velocity field $v_s(t)$ does not suffer these limitations. For any
hydrodynamic model considered, its correlation function can be
obtained using the LRT (see Sec.~\ref{sec3}). 
The resulting function is well behaved both at $t=0$, at which it
attains a constant value, and for $t \rightarrow \infty$, where it
vanishes in an integrable way (as its integral is finite and
corresponds to the particle diffusivity).

Most importantly, the stochastic velocity field $v_s(t)$ has a
clear physical meaning, as it corresponds to the stochastic velocity
of the particle in the absence of any external forcing or
perturbations. 
Consequently, $v_s(t)$
is amenable, in principle, to a direct experimental measurement from
the trajectory data of a free Brownian particle (which is not the case
of the thermal force $ R(t)$). However, it should be observed that in
real experiments the use of an external forcing, in the form of a
potential, e.g., deriving from the effects of an optical trap, is
always needed in order to localize particle motion in a given region
of the fluid domain, enabling the measurement of its position with
optical techniques.  In this case, the correlation functions of
$v_s(t)$ can be directly obtained experimentally by reducing
progressively the spring constant of the trap 
{(see Sec.~\ref{sec6}).}

It follows from the above discussion that 
introducing $v_s(t)$
as the descriptor of thermal fluctuations yields a different approach
to FD theory.  In classical statistical mechanics FD relations,
expressed by eqs.~(\ref{eq1_2})-(\ref{eq1_3}), are aimed at (i)
connecting velocity fluctuations to the properties of the hydrodynamic
response function (eq.~(\ref{eq1_2})); and (ii) characterizing the
thermal force $R(t)$ in terms of the hydrodynamic response function
(eq. ~(\ref{eq1_3})).  The action of hydrodynamic/thermal fluctuations
has thus been split into two contributions: dissipation (i.e.,
$h(t)$), and fluctuations (i.e., $R(t)$), and eq.~(\ref{eq1_3}) is
representative of the link connecting them. While this approach works
extremely well for eq.~(\ref{eq1_1}) 
{(see also Appendix~A)},
problems
arise in the presence of fluid inertial contributions
(Appendix~B). The presence
of hydrodynamic inertia makes this classical dichotomic description
blurred, in the meaning that fluid-inertial contributions cannot be
ascribed neither to pure dissipation nor to fluctuations. The present
theory based on the stochastic hydrodynamic velocity field formalizes
this concept, indicating that the separation between fluctuation and
dissipation in the description of thermal motion is, essentially, an
epistemic approach grounded on a model perspective of splitting the
various forces acting on the particle. Such a splitting derives from
comprehensible reasons, as it stems from the results on linear
hydrodynamics in the Stokes or in the time-dependent Stokes regime
\cite{kim} to which the thermal fluctuations are added as a further
contribution. The theory of Stochastic Hydrodynamics \cite{stocahydro}
provides a further interpretation of this result. Nonetheless, this is
still a model-based assuption in which the fluctuational term ${\bf
  R}(t)$ is added by physical necessity (as in the original paper by
Langevin \cite{langevin}). The stochastic velocity approach not only
resolves practical problems, such as those discussed above in
connection with the Basset force and further in the next section, but
treats the fluctuations in a unitary perspective, separating the role
of thermal fluctuations from the hydrodynamic effects. As a
consequence, given $v_s(t)$, no additional FD relations are needed for
describing thermal effects. It puts stochasticity at the center of the
physical focus, and asks for a physical interpretation of it.  In the
next two sections, a variety of examples are discussed in which
different models for the stochastic fluctuations, still possessing the
same velocity autocorrelation function, provide different, and
experimentally measurable, predictions.  Therein the formulation of
the stochastic velocity approach is finalized for catalyzing new
interest in this direction.

\section{Stochastic representation of hydrodynamic/thermal velocity 
 fluctuations}
\label{sec6}

The velocity splitting approach 
shifts the stochastic description of particle motion from the characterization of
the fluctuational force ${\bf R}(t)$ to the representation of the
comprehensive hydrodynamic/thermal stochastic velocity field ${\bf
  v}_s(t;{\bf x})$.
This naturally provides
a different setting for representing the stochastic velocity field in
terms of elementary stochastic processes.  This aspect is fundamental
in order to develop accurate stochastic Lagrangian descriptions
(Langevin equations) for particle motion.

In the early days of Einstein-Langevin investigations of Brownian
motion, the description of thermal fluctuations made use of stochastic
processes possessing no memory, i.e., characterized by an impulsive
correlation function. The use of Wiener processes $w(t)$, and of their
distributional derivatives $\xi(t)=dw(t)/d t$ (white noise), is the
natural and most convenient choice related to this level of
approximation.  This led to equations of motion of the form
\begin{subequations}
\begin{align} 
\dot{x}(t) & = v(t)  \\
m \dot{v}(t)  & = - \eta v(t) + \sqrt{2 k_B T \eta} \, \xi(t)\:,
\end{align}
\label{eq5_1}
\end{subequations}
where $\langle \xi(t^\prime) \xi(t) \rangle = \delta(t-t^\prime)$.
The use of $\delta$-correlated stochastic processes was adequate
to this level of approximation representing the physical phenomenology
of Brownian motion \cite{raizen2}, essentially because the
hydrodynamic interactions were considered to be instantaneous, as
expressed exclusively by the Stokesian drag. Reinterpreted in the light
of the velocity decomposition developed above, the statistical
properties of particle motion defined by eqs.~(\ref{eq5_1}) are
equally well predicted by the kinematic model
\begin{equation}
\dot{x}(t) = v_s(t)\:,
\label{eq5_2}
\end{equation}
 where $v_s(t)$ is any stochastic process possessing zero mean and
 exponential correlation function $\langle v_s(t) v_s(0) \rangle =
 \langle v^2 \rangle e^{-\eta t/m}$. We remark that $v_s(t)$
 represents the equilibrium velocity fluctuations, hence
 eq.~(\ref{eq5_2}) should not be confused with the classical
 overdamped approximation, valid in the limit $m \rightarrow 0$, as
 $v_s(t)$ in the present case is not $\delta$-correlated.  For
 instance, one could choose for $v_s(t)$ the Poisson-Kac process
 \cite{kac}
\begin{equation}
v_s(t) = b_0 (-1)^{\chi(t,\lambda)}\:,
\label{eq5_3}
\end{equation}
where $\chi(t,\lambda)$ is a Poisson counting process
characterized by the transition rate $\lambda>0$. Since
$\left \langle (-1)^{\chi(t,\lambda)} (-1)^{\chi(0,\lambda)} \right
\rangle = e^{-2 \lambda t}$,
the Poisson-Kac process 
fits the physical requirements, provided that $b_0^2 = \langle v^2 \rangle=k_BT/m$ and
$\lambda=\eta/2 m$.  This approach has been applied successfully even
in the presence of potentials \cite{venditti}. There is, however, an
important {\em caveat}.  While the choice eq.~(\ref{eq5_3}), applied
to the kinematics eq.~(\ref{eq5_2}), reproduces correctly all the
statistical properties of particle diffusional dynamics defined by
eqs.~(\ref{eq5_1}), it fails for describing the equilibrium velocity
probability density function. This is so, because eq.~(\ref{eq5_3})
represents a one-velocity model, characterized by a single velocity
value $b_0$, which determines an impulsive probability density
function for the velocity $v_s$, $p_{v}(v_s)=
[\delta(v_s+b_0)+\delta(v_s-b_0)]/2$.

\begin{figure}
\includegraphics[width=15cm]{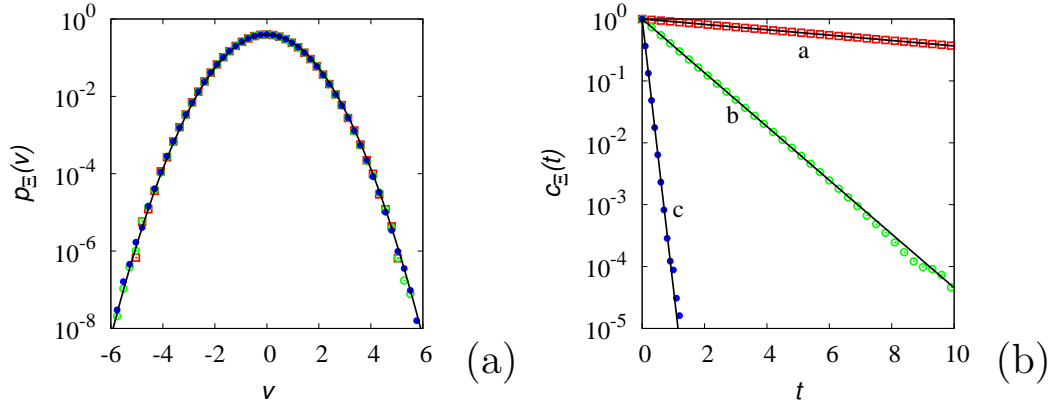}
\caption{Statistical properties of the process $\Xi_g(t,\lambda;v)$
  with $g(v)= e^{-v^2/2}/\sqrt{2 \pi}$.  Panel (a): Equilibrium
  probability density function $p(v)$ for
  $\Xi_g(t,\lambda;v)$. Symbols corresponds to the results of
  stochastic simulations at different values of
  $\lambda=10^{-1},\,1,1,10$. The solid line represents the normal
  probability density $g(v)$.  Panel (b): Correlation function
  $c_\Xi(t)$ vs.\ $t$.  Symbols corresponds to the results of
  stochastic simulations, lines to the exponential functions
  $c_\Xi(t)=e^{-\lambda t}$, eq.~(\ref{eq5_9}) since $\sigma_v^2=1$.
  Line(a) and ($\square$): $\lambda=10^{-1}$; line (b) and ($\circ$):
  $\lambda=1$, line (c) and ($\bullet$): $\lambda=10$.}
\label{Fig3}
\end{figure}

But in point of fact, this problem has a simple solution, as in all
the cases where an accurate reproduction of the velocity statistics is
required a Generalized Poisson-Kac process $\Xi_g(t,\lambda;v)$,
possessing velocity as a continuous transitional variable
\cite{giona2,rainer}, can be used instead of the conventional
dichotomous Poisson-Kac process (\ref{eq5_3}),
\begin{equation}
v_s(t) = \Xi_g(t,\lambda;v)\:.
\label{eq5_5}
\end{equation}
The process $\Xi_g(t,\lambda;v)$ is continuously parametrized with
respect to the velocity $v \in {\mathbb R}$, possesses an exponential
statistics of transition times specified by the transition rate
$\lambda$, and is such that the probability density function for $v$
would be any equilibrium function $g(v)$, for instance, the Maxwellian
distribution $g(v)=A e^{-mv_s^2/2 k_B T}$, where $A$ is the
normalization constant.  The statistical characterization of
$\Xi_g(t,\lambda;v)$ involves the probability density $P_\Xi(v,t)$, $v
\in {\mathbb R}$,
\begin{equation}
P_\Xi(v^\prime,t) d v^\prime = \mbox{Prob}[  \, \Xi_g(t,\lambda;v) \in (v^\prime,v^\prime+dv^\prime) \, ]\:,
\label{eq5_6}
\end{equation}
that satisfies the balance equation
\begin{equation}
\frac{\partial P_\Xi(v,t)}{\partial t}= - \lambda \, P_\Xi(v,t)
+ \lambda \, g(v) \int_{-\infty}^\infty P_\Xi(v^\prime,t) d v^\prime\:.
\label{eq5_7}
\end{equation}
Consequently, the statististical properties of the process $X(t)$ (we
use the notation $X(t)$ for the process and $x(t)$ for a realization
of it), defined by the kinematic eq.~(\ref{eq5_2}), are described by
the probability density $p(x,v,t)$ satisfying the linear Boltzmann
equation
\begin{equation}
\frac{\partial p(x,v,t)}{\partial t}= - v \frac{\partial p(x,v,t)}{\partial x}
-\lambda \, p(x,v,t) + \lambda \, g(v)   \int_{-\infty}^\infty p(x,v^\prime,t)\:.
\, d v^\prime\:.
\label{eq5_8}
\end{equation}
It is important to observe that the simulation of the process
$\Xi_g(t,\lambda;v)$ is as simple as the Poisson-Kac process
$(-1)^{\chi(t,\lambda)}$ since, at any transition time $\tau$, whose
statistics is defined by the exponential density $p_\tau(\tau)=\lambda
e^{-\lambda \, \tau}$, a new velocity variable is selected,
independently of the previous one, from the equilibrium distribution
$g(v)$.  Because of this property, the correlation function $c_\Xi(t)=
\langle \, \Xi_g(t,\lambda;v) \, \Xi_g(0,\lambda;v)\, \rangle$ is
given by
\begin{equation}
C_\Xi(t) = \sigma_v^2 e^{-\lambda t} \, , \quad
\sigma_v^2 = \int_{-\infty}^\infty v^2 \, g(v) \, d v
\label{eq5_9}
\end{equation}
independently of the functional form of the velocity probability
density $g(v)$.  
Figure~\ref{Fig3} compares the velocity
probability density function and the correlation function obtained
from stochastic simulations of the process $\Xi_g(t,\lambda;v)$ for the
case where $g(v)$ is the normal distribution for different values of
$\lambda$.  Simulations involve an ensemble of $10^8$ realizations.
The above example requires a further comment.  Generalized Poisson-Kac
processes have been developed in order to generate stochastic dynamics
characterized by bounded propagation velocity \cite{giona1}.  The
choice of a Gaussian probability density $g(v)$ for the statistics of
$\Xi_g(t,\lambda;v)$ is conceptually in contradiction with this
founding principle. The mathematical occurrence of the Gaussian
distribution can be justified by invoking the Central Limit Theorem,
and therefore it represents a long-term asymptotics.  As regards
particle velocities, the application of the Central Limit Theorem is
physically limited by relativistic constraints, due to the fact that
for large velocities the assumption of independence among the velocity
entries fails. The Maxwellian distribution thus represents an
excellent approximation of the relativistic J\"uttner distribution
\cite{juttner,gionaj} that, in the low-velocity/low-temperature limit
(with respect to the speed of light {\em in vacuo}), is practically
indistinguishable from its relativistic counterpart.  It follows from
the above reasoning that if we require the process $X(t)$ (i.e.\ the
particle position) to possess a bounded propagation velocity, the
process $\Xi_g(t,\lambda;v)$ defining $v_s(t)$ via eq.~(\ref{eq5_5})
should be constructed, e.g., by using a truncated Maxwellian
distribution, i.e.,
\begin{equation}
g(v)=
\left \{
\begin{array}{cc}
A e^{-m v^2/2 k_B T} &  \;\;  v \in (-v_{\rm max},v_{\rm max}) \\
0    & \mbox{otherwise}
\end{array}
\right .
\label{eq5_10}
\end{equation}
with $v_{\rm max} \gg \sqrt{k_B T/m}$ but still bounded, in order to
fulfill the Maxwellian equilibrium distribution in its physical range
of validity, but still satisfying the requirement of bounded
propagation.

The above construction applies {\em a fortiori} for particle dynamics
accounting for inertial hydrodynamic effects.  Considering the case
analyzed in Sec.~\ref{sec4}, due to the occurrence of asymptotic
power-law tails in the velocity correlation function, the natural
candidates for modeling $v_s(t)$ are L\'evy Walks
\cite{zaburdaev,fedotov}.  The classical L\'evy Walk model is
characterized by a transition rate $\lambda(\tau)$ depending on the
transition age $\tau$ (the transition age corresponds to the time
elapsed from the latest transition) in the form of \cite{fedotov}
\begin{equation}
\lambda(\tau) = \frac{\xi}{1+\tau/\tau_0}
\label{eq5_11}
\end{equation}
with $\xi>1$ and $\tau_0>0$.  The equilibrium correlation function of
this process fulfills the long-term scaling \cite{zaburdaev}
\begin{equation}
\langle v_s(t) v_s(0) \rangle \sim t^{-(\xi-1)}\:.
\label{eq5_12}
\end{equation}
 For the problem considered in Sec.~\ref{sec4}, the
$t^{-3/2}$-hydrodynamic scaling is matched by taking $\xi=5/2$.  The
same approach, applied above in the case of Poisson-Kac processes to
obtain a continuous distribution of velocities, can be {\em verbatim}
enforced in order to define a L\'evy Walk $\Lambda_g(t,\xi,\tau_0;v)$,
characterized by the transition rate eq.~(\ref{eq5_11}), by a
continuous probability density function $g(v)$ and by the correlation
properties pertaining to the corresponding one-velocity L\'evy Walk
model.
 
The above generalizations, either for Poisson-Kac, Generalized
Poisson-Kac processes or L\'evy Walks, have been analyzed in a broader
stochastic framework in \cite{rainer}, introducing the class of
Extended Poisson-Kac (EPK) processes that subsume the family of
stochastic processes possessing Markov and semi-Markov transition
mechanisms and arbitrary parametrization with respect to the
transitional variables.  It follows from the above analysis that the
velocity splitting approach 
finds
in EPK processes the natural and computationally simple candidates for
expressing $ v_s(t)$ (or $ v_s(t, x)$). This means that $v_s(t)$ can
be expressed with arbitrary accuracy by means of a linear combination
of independent EPK processes,
\begin{equation}
v_s(t)= \sum_{h=1}^{N_\Lambda}  a_h \,  \Lambda_g(t,\xi_h,\tau_{0,h};v)
+ \sum_{k=1}^{N_\Xi} b_k \, \Xi_g(t,\lambda_k;v)\:,
\label{eq5_13}
\end{equation}
where the number of processes $N_\Lambda$, $N_\Xi$, the process
parameters $\xi_h$, $\tau_{0,h}$, $\lambda_k$, and the expansion
coefficients $a_h$, and $b_k$ should be optimized with respect to the
correlation function $\langle v_s(t) v(0) \rangle$.  The details of
the expansion eq.~(\ref{eq5_13}) , and of parameter optimization, are
of little interest in the present analysis, and they will be developed
elsewhere in the light of specific hydrodynamic applications.  But the
application of eq.~(\ref{eq5_13}) to Brownian motion and to problems
deriving from transport in microfluidic systems opens up interesting
and new research directions as briefly outlined in Sec.~\ref{sec7}.

\subsection{Correlations in continuous vs.\ dichotomous models}
\label{sec6_1}

\begin{figure}
\includegraphics[width=15cm]{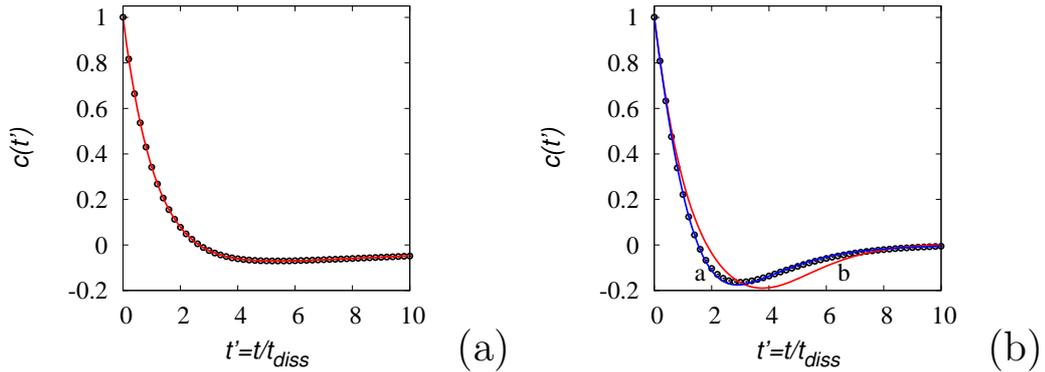}
\caption{Velocity autocorrelation function $c(t^\prime)$ vs
  $t^\prime=t/t_{\rm diss}$ associated with the stochastic dynamics
  eq.~(\ref{eqx_2}).  Symbols ($\bullet$) corresponds to the results
  of the stochastic simulation of eq.~(\ref{eqx_2}) Panel (a) refers
  to a harmonic potential at $\alpha=0.1$.  The solid line represents
  the velocity autocorrelation function obtained from the velocity
  split model eq.~(\ref{eqx_4}) describing $u_s(t)$ by means of a
  dichotomous Poisson-Kac process.  Panel (b) refers to the case of a
  bistable potential described in the main text. Line (a) represents
  the velocity autocorrelation function obtained using a continuously
  parametrized EPK process $\Xi(t^\prime,1;v)$ possessing a Maxwellian
  velocity density function, line (b) the corresponding result
  adopting for $u_s(t^\prime)$ a dichotomous Poisson-Kac process.}
\label{Fig4}
\end{figure}

In order  to highlight the importance of the transition
from dichotomous Poisson-Kac processes to EPK processes possessing
a continuous parametrization with respect to velocity in the
characterization of the $v_s(t)$, let us consider
the case of a particle in a thermostated fluid environment
at constant temperature $T$, subjected to an external
potential $U(x)$. In the one-dimensional case, the
particle dynamics reads
\begin{eqnarray}
d x & =  & v \, d t \nonumber  \\
m \, d v & = & - \eta \, v \, dt - \partial_x U(x) \, dt +
\sqrt{2 k_B T \eta} \, d w(t)\:.
\label{eqx_1}
\end{eqnarray}
Introducing $t= T_c \, t^\prime$, $x= L_c \, y$, $v= V_c \, u$,
where $T_c, \, L_c,\,V_c$ are the characteristic time, length and velocity
scales, letting $U(x)=U_0 \, \overline{U}(y)|_{y=x/L_c}$, and
assuming $T_c=m/\eta=t_{\rm diss}$, $V_c= \sqrt{k_B T/m}$, $L_c=V_c T_c$,
eqs. (\ref{eqx_1}) take the non-dimensional form
\begin{eqnarray}
d y & =  & u \, d t^\prime \nonumber  \\
 d u & = & -  u \, dt^\prime - \alpha \, \partial_y \overline{U}(y) \, dt^\prime +
\sqrt{2} \, d w(t^\prime)\:,
\label{eqx_2}
\end{eqnarray}
where $\alpha= U_0/m V_c^2$. Consider two cases. The first case
is represented by a harmonic potential $U(x)=k_s x^2/2$, so that $U_0=k_S L_c^2$.
In this case, $\alpha=U_0/m V_c^2= k_s L_c^2/m V_c^2=k_s T_c^2/m$, i.e.,
\begin{equation}
\alpha= \frac{k_s \, m}{\eta^2}= \frac{t_{\rm diss}}{t_k}\:,
\label{eqx_3}
\end{equation}
where $t_k=\eta/k_s$ is the characteristic time associated with the
coupling of the harmonic potential and frictional dissipation, and
$\partial_y \overline{U}(y)=y$.  This case represents the dynamics of
a Brownian particle (neglecting fluid inertial effects) in an optical
trap, and physically reasonable values for $\alpha$ range from
$10^{-2}$ to $10^{-1}$ \cite{raizenrev1}.  Specifically,
fluid-inertial effects are negligible if the fluid is a gas
\cite{raizenrev1}.  Figure \ref{Fig4} panel (a) depicts the comparison
of the velocity autocorrelation function obtained from the direct
simulation of eq.~(\ref{eqx_1}), for $\alpha=0.1$, using $N=10^6$
realizations of the process, with the one from the velocity split
representation
\begin{eqnarray}
d y & =  & (u_d + u_s(t)) \, d t^\prime \nonumber  \\
 d u_d & = & -  u_d \, dt^\prime - \alpha \, \partial_y \overline{U}(y) \, dt^\prime \:,
\label{eqx_4}
\end{eqnarray}
where the nondimensional stochastic velocity $u_s(t)$ is
represented by a Poisson-Kac dichotonous stochastic process.
In this case, the dichotomous model correctly reproduces the
behaviour of the velocity  autocorrelation function, as predicted
by LRT.

\begin{figure}
\includegraphics[width=10cm]{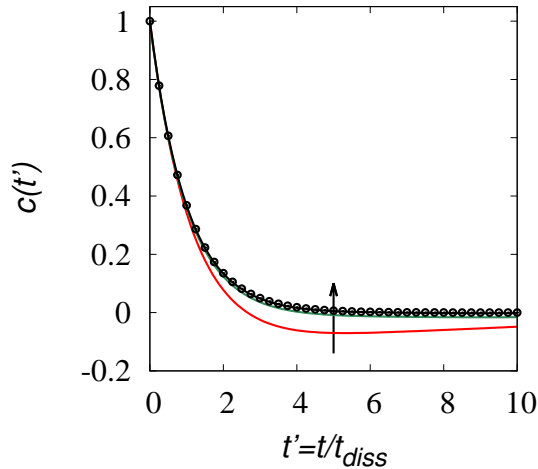}
\caption{Velocity autocorrelation function $c(t^\prime)$
  vs.\ $t^\prime=t/t_{\rm diss}$ associated with the stochastic
  dynamics eq.~(\ref{eqx_2}) in the presence of a harmonic potential
  for different values of $\alpha$, i.e., of the nondimensional spring
  constant, $\alpha=10^{-1}, \, 10^{-2}, \, 10^{-3}$.  The arrow
  indicates decreasing values of $\alpha$.  Symbols ($\circ$)
  correspond to the free-particle velocity autocorrelation function,
  $c(t^\prime)=e^{-t^\prime}$.}
\label{Fig4a}
\end{figure}

This model provides clear evidence for what was stated in
Sec.~\ref{sec5}, and specifically that the velocity autocorrelation
function $\langle v_s(t) \, v_s(0) \rangle$ can be measured in
physically realizable experiments.  Indeed, still keeping the Brownian
particle confined in a trap (and therefore under the influence of a
harmonic potential), by decreasing the trap spring constant and
measuring the corresponding particle velocity autocorrelation function
$\langle v(t) \, v(0) \rangle$, one obtains direct and accurate
experimental measurements of the free-particle velocity
autocorrelation function, i.e., of $\langle v_s(t) v_s(0) \rangle$
(see Fig.~\ref{Fig4a}). For values of the nondimensional parameter
$\alpha$ less or equal to $10^{-2}$, the velocity autocorrelation
function of the trapped particle represents a sufficiently accurate
representation for that of the free particle, and thus of the
autocorrelation function of $u_s(t)$.  In experiments involving
trapped micrometric particles, the values of $\alpha$ are usually
small enough to operate in this limit \cite{raizenrev1}.
Next, consider a nonlinear potential, such as the bistable potential
expressed in nondimensional form by $\overline{U}(y)=y^4/4-y^2/2$.
Figure~\ref{Fig4} panel (b) depicts the temporal behaviour of the
velocity autocorrelation function obtained from the stochastic
simulation of eq.~(\ref{eqx_2}) at $\alpha=0.1$, and for the velocity
split model eq.~(\ref{eqx_4}) in the case $u_s(t)$ is an EPK process
continuously parametrized with respect to the velocity and possessing
a Maxwellian distribution (line a) and a dichotomous Poisson-Kac
process (line b).  Owing to the nonlinearity of the model, the
importance of a continuous distribution of velocity is evident.  While
the EPK process $\Xi(t,\lambda;v)$ accurately reproduces the velocity
autocorrelation function, this is not the case for the dichotomous
Poisson-Kac counterpart. This indicates that when moving towards
nonlinear models, in the present case expressed by a non-quadratic
potential, the fine structure of the velocity fluctuation becomes
important.

\subsection{On the fine structure of the thermal fluctuations}
\label{sec6_2}

In Kubo FD theory, and in transport modeling, the velocity
autocorrelation function is the statistical quantity used to
characterize thermal fluctuations.  This is inherent to transport
theory, as the integral over time of the velocity autocorrelation
function returns the diffusion coefficient (Green-Kubo theorem).  More
generally, the whole architecture of statistical physics of
non-equilibrium phenomena is grounded on the autocorrelation functions
of fluctuating fluxes and forces \cite{zwanzx,kirkx}.

The recent advances in the experimental analysis of Brownian
motion at short time scales have revealed the possibility of
characterizing Brownian fluctuations beyond the limit that ``Einstein
deemed possible'' \cite{raizen2}. The experimental investigation of
Brownian motion can nowadays provide a deeper understanding of the
physical meaning of thermal noise, of the interaction between
hydrodynamic and thermodynamic properties, and possibly of the
macroscopic dissipative effects of quantum fluctuations, either at the
level of radiative interactions with molecules and particles
(involving photon exchange) or at the field level involving zero-point
fluctuations. In this perspective, the analysis of the fine properties
of the stochastic velocity field $v_s(t)$ and of its influence on
macroscopic and experimentally measurable quantities becomes central.

\begin{figure}
\includegraphics[width=15cm]{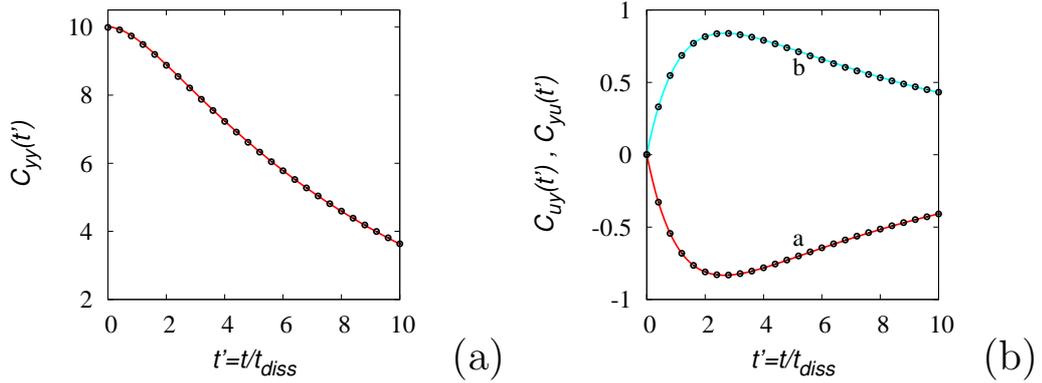}
\caption{Second-order correlation functions associated with the
  stochastic dynamics eq.~(\ref{eqx_2}) in the presence of a harmonic
  potential with $\alpha=0.1$.  Solid lines represent the correlation
  functions obtained from the velocity split model eq.~(\ref{eqx_4}),
  using for $u_s(t^\prime)$ a continuously parametrized EPK process
  $\Xi(t^\prime,1;v)$ possessing a Maxwellian velocity probability
  density function. Panel (a): $C_{yy}(t^\prime)$ vs.\ $t^\prime$.
  Panel (b): $C_{uy}(t^\prime)$ (line a) and $C_{yu}(t^\prime)$ (line
  b) vs.\ $t^\prime$. Dots correspond to correlation functions
    obtained from solving eq.~(\ref{eqx_2}).}
\label{Fig4b}
\end{figure}

Consider again the case of a Brownian particle in a trap (i.e., where
$U(x)$ is a harmonic potential) described by means of
eq.~(\ref{eqx_2}). Experimentally, eq.~(\ref{eqx_2}) corresponds to
particle motion in a gaseous environment at thermal equilibrium for
which the fluid-inertial interactions are negligible.  Consider the
other second-order correlation functions, namely $C_{yy}(t^\prime)=
\langle y(t^\prime) \, y(0) \rangle$, $C_{yu}(t^\prime)=\langle
y(t^\prime) \, u(0) \rangle$, and $C_{uy}(t^\prime)= \langle
u(t^\prime) \, y(0) \rangle$.  Due to the linearity of the dynamics,
any stochastic representation of $u_s(t^\prime)$, entering
eq.~(\ref{eqx_4}) and possessing the correct exponential decay of the
velocity autocorrelation function, would provide the same temporal
behavior for all the second-order correlation functions.  This
phenomenon is depicted in Fig.~\ref{Fig4b}, where the correlation
function obtained from the solutions of eq.~(\ref{eqx_2}) are compared
with the corresponding ones obtained from eq.~(\ref{eqx_4}) adopting
for $u_s(t^\prime)$ an EPK model possessing a Maxwellian probability
density function as described and used in the previous paragraph. 
{An analogous result could be obtained adopting for $u_s(t^\prime)$ a
dichotomous one-velocity Poisson-Kac model.}

\begin{figure}
\includegraphics[width=10cm]{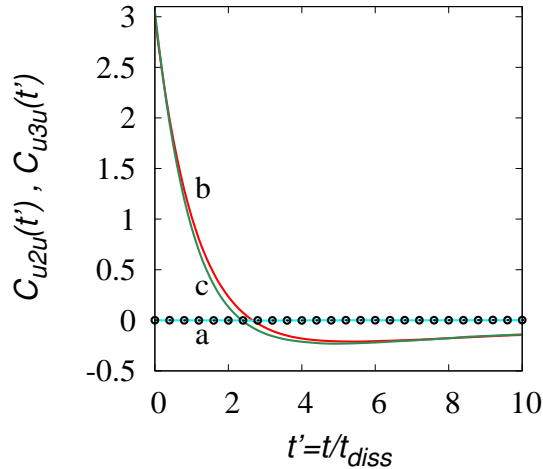}
\caption{Higher-order velocity autocorrelation functions associated
  with the stochastic dynamics eq.~(\ref{eqx_2}) in the presence of a
  harmonic potential with parameter $\alpha=0.1$.  Line (a) and line
  (b) refer to $C_{u2u}(t^\prime)$ and $C_{u3u}(t^\prime)$,
  respectively, obtained from the solutions of
  eq.~(\ref{eqx_2}). Symbols ($\circ$) and line (c) refer to
  $C_{u2u}(t^\prime)$ and $C_{u3u}(t^\prime)$, respectively, obtained
  from the solutions of eq.~(\ref{eqx_4}) using for $u_s(t^\prime)$ a
  continuously parametrized EPK process $\Xi(t^\prime,1;v)$ possessing
  a Maxwellian velocity probability density function.}
\label{Fig4c}
\end{figure}

\begin{figure}
\includegraphics[width=15cm]{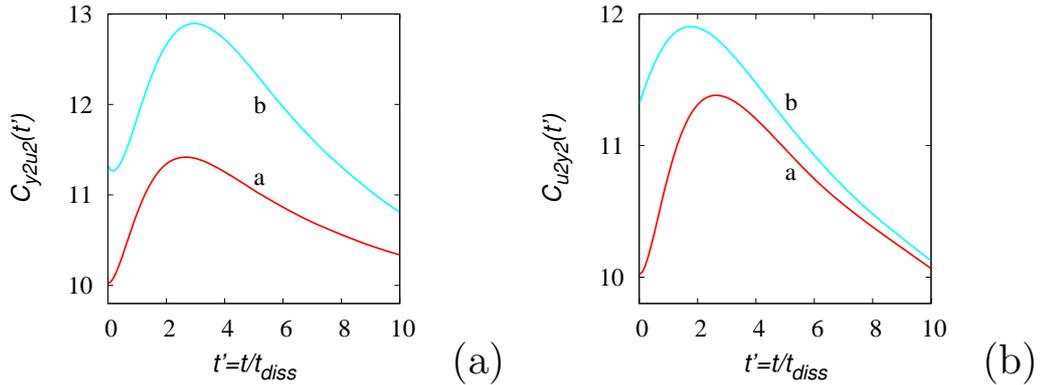}
\caption{Higher-order mixed correlation functions for a particle in a
  harmonic potential.  Panel (a) depicts $C_{y2u2}(t^\prime)$
  vs.\ $t^\prime$, panel (b) $C_{u2y2}(t^\prime)$ vs.\ $t^\prime$.
  Lines (a) refer to the solutions of eq.  (\ref{eqx_2}), lines (b) to
  the solutions of eq.~(\ref{eqx_4}) using for $u_s(t^\prime)$ a
  continuously parametrized EPK process $\Xi(t^\prime,1;v)$ possessing
  a Maxwellian velocity probability density function.}
\label{Fig4d}
\end{figure}

Recent experimental studies suggest that velocity fluctuations of
Brownian particles both in gases and liquids could be characterized by
a much more regular (smooth) behavior than that predicted by the
classical Einsteinian theory eq.~(\ref{eqx_2}), based on Wiener
processes as a model of thermal fluctuations \cite{exp3,raizen2}. The
experimental assessment of the regularity of thermal fluctuations
represents an important issue in statistical physics, as it would
suggest the necessity of considering more regular stochastic processes
for the modeling of particle transport. Beside a regularity analysis
of Brownian trajectories, clear experimental evidence of these
properties may come from the estimate of higher-order correlation
functions.  Figure~\ref{Fig4c} depicts the comparison of the third and
fourth-order velocity autocorrelation functions $C_{u2u}(t^\prime)=
\langle u^2(t^\prime) \, u(0) \rangle$, $C_{u3u}(t^\prime) = \langle
u^3(t^\prime) \, u(0) \rangle$, obtained from the classical
Wiener-based model eq.~(\ref{eqx_2}), and from eq.~(\ref{eqx_4}) in
the presence of a piecewise smooth description of the stochastic
velocity field $u_s(t^\prime)$ via the EPK model described in the
previous section, possessing a Maxwellian velocity distribution
function. While $C_{u2u}(t^\prime)=0$ in both cases, a slight
difference between the two descriptions of the thermal fluctuations is
observed in the fourth-order autocorrelation function
$C_{u3u}(t^\prime)$, although this difference is too small to be an
object of experimental scrutiny. Conversely, a significant, and
experimentally verifiable discrepancy between the Wiener description
and and a smoother regular descriptions of thermal velocity
fluctuations characterizes higher-order (fourth-order) mixed
correlation functions, such as $C_{y2v2}(t^\prime)=\langle
y^2(t^\prime) \, u^2(0) \rangle$ or $C_{v2y2}(t^\prime)= \langle
u^2(t^\prime) \, y^2(0) \rangle$, depicted in Fig.~\ref{Fig4d}.

We may therefore conclude from these qualitative observations that the
experimental analysis of higher-order correlation functions of
position and velocity variables may provide a way, not only for
obtaining a finer statistical characterization of the stochastic
velocity field $v_s(t)$, but also for verifying quantitatively, via
stable and accurate experimental measurements, the validity of the
Einsteinian paradigm for thermal fluctuations based on almost
everywhere singular stochastic processes (Wiener process), and to
assess experimentally the regularity of thermal fluctuations. This
would close the circle on the properties of Brownian motion that at
times of Einstein would be unthinkable to assess experimentally
\cite{raizen2}.

\section{Concluding remarks}
\label{sec8}
The classical FD formalism involves a decomposition of
hydrodynamic/thermal effects into a dissipative contribution and a
purely stochastic fluctuational term, including eventually
fluid-inertial back action.  The separation of fluid inertial effects
(the Basset force) from the thermal contribution ${\bf R}(t)$ is
essentially a technical simplification not related to the physics of
the problem, which conversely indicates a highly correlated motion of
the particle and of its nearby fluid elements \cite{darwin}. An
alternative to this approach is to consider all these
hydrodynamic/thermal effects as a single fluctuating field possessing
prescribed statistical and correlation properties.  This is the key
idea of the velocity splitting approach that we developed here, which
{provides a comprehensive description of these interactions in
  terms of the stochastic process ${\bf v}_s(t)$, or the stochastic
  field ${\bf v}_s(t;{\bf x})$.}  The correlation properties of ${\bf
  v}_s(t,{\bf x})$ can be derived from linear response theory even in
those cases - as for particle transport in confined systems - where
both dissipative and inertial effects depend on the particle position,
by considering the conditional correlation functions ${\bf c}^{(k)}(t
\, | \, {\bf x})$.

{Within this framework,} 
the stochatic force ${\bf R}(t)$ is no longer necessary, 
as it is encompassed, together with all the
(dissipative/inertial) hydrodynamic thermal interations, in the
stochastic velocity field ${\bf v}_s(t;{\bf x})$. This field is
a directly measurable quantity in short-time Brownian motion
experiments, which is not the case for the thermal force ${\bf R}(t)$
in hydrodynamic problems involving not only friction but also fluid
inertial effects. In the latter case we have shown that, apart from
formal results, it is difficult to obtain a computationally operative
definition of ${\bf R}(t)$ in terms of elementary stochastic processes
to be used in the numerical simulation of the corresponding Langevin
equations.  

For modeling ${\bf v}_s(t;{\bf x})$, the class of EPK processes
emerges as a natural choice, due to their simplicity in the
implementation and due to the typical exponential/power-law behaviour
of their correlation functions. The decomposition of a generic
stochastic velocity field ${\bf v}_s(t;{\bf x})$ into a family of EPK
process is an interesting computational problem that will be addressed
in forthcoming works.

Pursuing this alternative approach, and comparing it with the results
derived from classical Wiener-based Langevin equations, in several examples we have shown that
\begin{itemize}
\item the sole velocity autocorrelation function is not sufficient to
  describe correctly Brownian motion and micrometric particle dynamics
  in the presence of nonlinear potentials, which points at a more
  detailed characterization (both theoretically and experimentally) of
  thermal fluctuations;
\item higher-order correlation functions, or correlation functionals, can be
used as probes for determining the fine structure of the stochastic
fluctuations.
\end{itemize}

{
%
In this manuscript, 
we have mainly considered the description of
${\bf v}_s(t)$ in the free space, where 
no external deterministic accelerations ${\bf a}({\bf x})$ or
position-dependent hydrodynamic effects are present. 
Whenever present, these determine the explicit
dependence of the linear functionals $L_d$ and $L_i$ on ${\bf x}$.
The presence of external position-dependent fields generating ${\bf
  a}({\bf x})$ does not present any conceptual issue, because the
statistical structure of the hydrodynamic/thermal stochastic velocity
${\bf v}_s(t)$ is independent of ${\bf a}({\bf x})$ 
(see Subsec.~\ref{sec6_1}).
In the case of position-dependent hydrodynamic
interactions, which occurs for particles in confined geometries, such as 
in microchannels of transversal length scale comparable to the particle diameter, 
the statistical properties of the stochastic velocity field ${\bf
  v}_s(t;{\bf x})$ can be derived, for fixed ${\bf x}$, from the
conditional normalized correlation functions ${\bf c}^{(k)}(t \, |
{\bf x})$ satisfying eqs.~(\ref{eq3_7}) (see Sec.~\ref{sec3}).}
{In this setting, it is no longer possible to
derive {\em a priori} the velocity correlation function, or the
statistical characterisation of the spatial particle distribution,
even in the long-term limit (if an asymptotic equilibrium distribution
emerges in the system). 
At present, these 
can be obtained solely from
the direct numerical simulations of the equations. 
The theoretical prediction of these properties
represents a major technical issue, 
whose solution would necessarily rely on the extension of LR theory to
the nonlinear case. 
The outline of a specific setup to be studied is further described in Appendix \ref{sec7}.}

The experimental characterization of the fine
structure of the stochastic velocity fields ${\bf v}_s(t;{\bf x})$
defines another important topic for future investigations that can be
addressed by measuring higher-order correlation functions. 
This analysis could potentially be used to check for the
reliability of the singular Wiener-based approach to characterize
thermal fluctuations at short time scales, which is in contrast to a
more smooth and piecewise regular description of equilibrium
fluctuations that seems to emerge from experiments \cite{raizen2}.

We conclude by observing that while the
classical decomposition of hydrodynamic/thermal interactions into
dissipative, inertial, and fluctuational forces follows naturally from
a reductionistic Newtonian approach of accounting for all the
``distinct'' forces acting on a material body, our formulation of the
statistical physical properties of a particle in a thermalized fluid,
based on the 
stochastic velocity field ${\bf
  v}_s(t;{\bf x})$, shifts the focus of the description on the
kinematic equation for the particle motion driven by the ``almost
deterministic'' velocities ${\bf v}_d(t)$, and by the stochastic
velocities ${\bf v}_s(t;{\bf x})$. This alternative decomposition
bears some analogies with the Aristotelian description of motion,
where ${\bf v}_d(t)$ expresses the ``violent motion'' and ${\bf
  v}_s(t;{\bf x})$ the ``natural motion'' in thermal systems at
equilibrium \cite{rovelli}.

\vspace*{0.5cm}

\appendix

\section{Modal decomposition of the thermal force - The case of purely dissipative hydrodynamics}

Consider the dissipative Langevin equation (\ref{eq1_1}). In this case, the Kubo FD theory
predicts for the correlation function of the stochastic forcing
\begin{equation}
\langle R(t) \, R(0) \rangle =  \, m \, k_B \, T h(t) \, , \quad t \geq 0\:,
\label{eq_app1}
\end{equation}
where we have assumed $\langle \xi(t) \, \xi(0) \rangle =\delta(t)$  for the distributional
derivative  $\xi=d w(t)/dt$ of a Wiener process.
The friction kernel $h(t)$ satisfies the property
\begin{equation}
h(t) \geq 0 \, , \qquad t \geq 0\:.
\label{eq_app2}
\end{equation}
Let us assume that it admit a modal representation of the form
\begin{equation}
h(t) = \frac{1}{m} \sum_{i=1}^\infty a_i \, e^{-\lambda_i \, t}\:,
\label{eq_app3}
\end{equation}
where $a_i, \,\lambda_i >0$, $\sum_{i=1}^\infty  a_i < \infty$. The stochastic forcing
$R(t)$ can be expressed  in the following form,
\begin{equation}
R(t)= \sum_{i=1}^\infty \sqrt{ k_B \, T \, b_i} \, \psi_{\lambda_i}(t)\:,
\label{eq_app4}
\end{equation}
where $\psi_{\lambda_i}(t)$, $i=1,\dots,n$ are stochastic processes, independent
of each other and possessing exponential correlation functions
\begin{equation}
\langle \psi_{\lambda_i}(t) \, \psi_{\lambda_j}(t^\prime) \rangle = \delta_{i,j}
\, e^{-\lambda_i \, |t-t^\prime|}\:,
\label{eq_app5}
\end{equation}
where the positive constants $b_i >0$ need to be determined. The correlation
function of $R(t)$ is given by
\begin{equation}
\langle R(t) \, R(0) \rangle =   k_B \, T \, 
\sum_{i=1}^\infty \sqrt{b_i} \sum_{j=1}^\infty \sqrt{b_j} \, \langle \psi_{\lambda_i}(t) \, \psi_{\lambda_j}(0) \rangle =  k_B \, T \, \sum_{i=1}^\infty b_i \, e^{-\lambda_i \, t}\:,
\label{eq_app6}
\end{equation}
and from eqs.~(\ref{eq_app1}), (\ref{eq_app3}) it follows that the
expansion coefficients $b_i$ are simply given by
\begin{equation}
b_i= a_i \, ,  \quad i=1,\dots \:.
\label{eq_app7}
\end{equation}
The processes $\psi_{\lambda_i}(t)$ characterized by the property
eq.~(\ref{eq_app5}) can be defined in many ways. For instance, they
can be chosen in the form of filtered Wiener processes,
\begin{equation}
d \psi_{\lambda_i}(t) = - \lambda_i \, \psi_{\lambda_i}(t) \, d t+ \sqrt{2 \lambda_i} \, d w_i(t)\:,
\label{eq_app8}
\end{equation}
where $d w_i(t)$ are the increments of independent one-dimensional
Wiener processes so that $\langle d w_i(t) \, d w_j(t) \rangle =
\delta_{i,j} \, d t$. Or they can be represented in terms of
Poisson-Kac processes,
\begin{equation}
\psi_{\lambda_i}(t)= (-1)^{\chi_i(t,\lambda_i/2)}\:,
\label{eqapp9}
\end{equation}
where $\chi_i(t,\lambda_i/2)$ are independent Poisson processes
characterized by the transition rate $\lambda_i/2$, $ \left \langle
(-1)^{\chi_i(t,\lambda_i/2)} \, (-1)^{\chi_j(t,\lambda_j/2)} \right
\rangle = \delta_{i,j}$. Or one may use independent EPK processes,
possessing exponential correlation functions and arbitrary velocity
probability density functions with zero mean and unit variance.

The same approach can be extended to a continuous representation. In
that case $h(t)$ can be represented in the form
\begin{equation}
h(t)= \frac{1}{m} \int_0^\infty a(\lambda) \, e^{-\lambda \, t} \, d t
\label{eqapp10}
\end{equation}
with $\alpha(\lambda) \geq 0$.  Here the thermal force can be
expressed as
\begin{equation}
R(t)= \sqrt{ k_B \, T} \, \int_0^\infty \sqrt{a(\lambda)} \, \psi_\lambda(t) \, d \lambda\:,
\label{eqapp11}
\end{equation}
where
\begin{equation}
\langle \psi_\lambda(t) \, \psi_\mu(t^\prime) \rangle = \delta(\lambda-\mu) \, 
e^{-\lambda \, |t-t^\prime|}\:.
\label{eqapp12}
\end{equation}
It follows from this analysis that the thermal forces admit simple and
compact representations in the presence of purely dissipative
hydrodynamic contributions, and that there are in principle infinitely
many elementary systems of stochastic processes $\psi_\lambda(t)$ that
can be used to represent $R(t)$ in a way consistent with the FD
theorem.\\

\section{Representation of thermal forces in the presence of fluid inertia}

In this Appendix the  application of the classical FD theorem in the presence
of fluid inertia is discussed. In this case, for large $\omega$ (see Sec.~\ref{sec5}),
\begin{equation}
\mbox{Re}[\zeta(\omega)] \sim \omega^{1/2}\:,
\label{eqbpp1}
\end{equation}
which implies that  $\langle R(t) \, R(0) \rangle \sim t^{-3/2}$ at short timescales.
Therefore, without loss of generality let us assume that
\begin{equation}
\langle R(t) \, R(0) \rangle = \frac{C}{t^{3/2}}\:,
\label{eqbpp2}
\end{equation}
where $C$ is a constant, for fixed $T$, the actual value of which is
inessential in the present analysis. Also, in this case $R(t)$ admits
a modal representation in the form of eq.~(\ref{eqapp11}), say
$R(t)=\int_0^\infty \sqrt{b(\lambda)} \, \psi_\lambda(t) \, d
\lambda$, where $b(\lambda)$ is the solution of the functional
equation
\begin{equation}
\int_0^\infty b(\lambda) \, e^{-\lambda \, t} \, d \lambda =
\frac{C}{t^{3/2}}
\label{eqbpp3}
\end{equation}
indicating that $b(\lambda)$ is the inverse Laplace transform of the
correlation function $\langle R(t) \, R(0) \rangle$ in which time $t$
plays the role of the Laplace variable.  Using known results of the
theory of Laplace transforms, eq.~(\ref{eqbpp3}) admits the solution
\begin{equation}
b(\lambda)= \frac{2 \, C}{\sqrt{\pi}} \, \sqrt{\lambda}\:.
\label{eqbpp4}
\end{equation}
Consider $\langle R^2(t) \rangle$. From the above relations, it follows that
\begin{equation}
\langle R^2(t) \rangle = \frac{4 \, C^2}{\pi} \int_0^\infty \lambda^{1/4} \, d \lambda
\int_0^\infty \mu^{1/4} \, \langle \psi_\lambda(t) \, \psi_\mu(t) \rangle \, d \mu
= \frac{4 \, C^2}{\pi} \int_0^\infty \lambda^{1/2} \, d \lambda = \infty\:.
\label{eqbpp5}
\end{equation}
Therefore, the second-order moment of the thermal force is unbounded,
which is indeed a highly singular and unpleasent property. One could
argue that this is also the case for the distributional derivative
$\xi(t)=d w(t)/dt$ of a Wiener process, modelling the thermal forces
in the presence of a purely instantaneous Stokesian friction.  In the
latter case, however, the infinitesimal increments
\begin{equation}
d F(t) = R(t) \, d t
\label{eqbpp6}
\end{equation}
over time $dt$ possess bounded second-order moments $\langle d F^2(t)
\rangle \sim d t$, proportional to $d t$.  It is therefore interesting
to evaluate this quantity, in the case of the process defined by
eq.~(\ref{eqbpp4}). Introduce the new process $q_\lambda(t)$, defined
by the relation
\begin{equation}
d q_\lambda(t)= \psi_\lambda(t) \, d t \,  \quad \Rightarrow \quad q_\lambda(t)=\int_0^t \psi_\lambda(\tau) \, d \tau
\label{eqbpp7}
\end{equation}
where, conventionally, $q_\lambda(0)=0$. Thus
\begin{equation}
d F(t)= \int_0^\infty \sqrt{b(\lambda)} \, d q_\lambda(t) \, d \lambda =
\int_0^\infty \sqrt{b(\lambda)} \int_t^{t+d t} \psi_\lambda(\tau) \, d \tau\:.
\label{eqbpp8}
\end{equation}
Therefore $\langle d F^2(t) \rangle$ can be evaluated to
\begin{eqnarray}
\langle d F^2(t) \rangle  & =  & \int_0^\infty \sqrt{b(\lambda)} \, d \lambda \int_0^\infty 
\sqrt{b(\mu)} \, d \mu \int_t^{t+ dt} d \tau \int_t^{t+dt} \langle \psi_\lambda(\tau)
\, \psi_\mu(\theta) \, d \theta \nonumber \\
& = & \int_0^\infty b(\lambda) \, d \lambda \int_t^{t+d t} d \tau \int_t^{t+ dt} 
e^{-\lambda | \tau-\theta|} \, d \theta
= 2  \,  \int_0^\infty b(\lambda) \,  d \lambda \int_t^{t+d t} d \tau \int_t^\tau e^{-\lambda (\tau-\theta)}
d \theta \nonumber \\
& = & 2  \, \int_0^\infty \frac{b(\lambda)}{\lambda} \, d \lambda 
\left [ \int_t^{t+ d t} d \tau - e^{\lambda  \, t} \int_t^{t+d t} e^{-\lambda \, \tau} \, d \tau
\right ] \nonumber \\
& = &
2 \,  \int_0^\infty \frac{b(\lambda)}{\lambda} \, d \lambda \, d t + 2 \, 
\int_0^\infty \frac{b(\lambda)}{\lambda^2} \,  \left (e^{-\lambda \, d t}-1  \right )
\, d \lambda = 2 \, \int_0^\infty b(\lambda) d \lambda \, d t^2\:.
\label{eqbpp9}
\end{eqnarray}
Consequently, due to the singular behavior of $b(\lambda)$ at
infinity, see eq.  (\ref{eqbpp4}), $\langle d F^2(t) \rangle =\infty$
for any $d t$. This essentially implies that either $R(t)$ or $d F(t)$
admit a purely formal modal representation in terms of elementary
stochastic processes which, hovewer, does not correspond to any
physically realizable stochastic evolution.

\section{Application to microfluidics and transport in confined geometries}
\label{sec7}

In this paper we have mainly considered the description of
${\bf v}_s(t)$ in the free space in the absence either of external
deterministic accelerations ${\bf a}({\bf x})$ or of
position-dependent hydrodynamic effects, determining the explicit
dependence of the linear functionals $L_d$ and $L_i$ on ${\bf x}$.
The presence of external position-dependent fields generating ${\bf
  a}({\bf x})$ does not present any further problem, as the
statistical structure of the hydrodynamic/thermal stochastic velocity
${\bf v}_s(t)$ is independent of ${\bf a}({\bf x})$.  This case has
been briefly addressed in Subsec.~\ref{sec6_1}, in order to highlight
the importance of a continuous velocity distribution for ${\bf
  v}_s(t)$.  In the case of position-dependent hydrodynamic
interactions, which occurs for particles in microchannels of
transversal lengthscale comparable to the particle diameter, the
statistical properties of the stochastic velocity field ${\bf
  v}_s(t;{\bf x})$ can be derived, for fixed ${\bf x}$, from the
conditional normalized correlation functions ${\bf c}^{(k)}(t \, |
{\bf x})$ defined in Sec.~\ref{sec3} and satisfying
eqs.~(\ref{eq3_7}).

What makes these problems peculiar with respect to the free space case
is that it is no longer possible, due to the generic nonlinear
dependence of ${\bf a}({\bf x})$, $L_d$ and $L_i$ on ${\bf x}$, to
derive {\em a priori} the velocity correlation function, nor the
statistical characterization of the spatial particle distribution,
even in the long-term limit (if an asymptotic equilibrium distribution
emerges in the system). This is the case whenever hydrodynamic
effects (convective fluxes) cope with potential contributions
(conservative forces) so that the resulting acceleration field ${\bf
  a}({\bf x})$ possesses a full Helmholtz decomposition,
\begin{equation}
{\bf a}({\bf x}) = \nabla \phi_a({\bf x})+ \nabla \times {\bf K}_a({\bf x})\:,
\label{eq6_1}
\end{equation}
where the scalar, $\phi_a({\bf x})$, and vector, ${\bf K}_a({\bf x})$,
potential are both different from zero.  In these cases, the
understanding of the emergent statistical properties relies on the
direct simulation of particle dynamics. In the velocity split
approach, this implies the simulation of the stochastic differential
equations
\begin{eqnarray} 
 \dot{\bf x}(t) &  = &{\bf v}_d(t)+{\bf v}_s(t,{\bf x}(t))  \nonumber \\
 \dot{\bf v}_d(t)  &= & 
 L_d[{\bf v}_d(t);{\bf x}(t)] + L_i[\dot{\bf v}_d(t);{\bf x}(t)] + 
{\bf a}({\bf x}(t))\:,
\label{eq6_2}
 \end{eqnarray}
where the choice of the simplest and most efficient representation of
the stochastic field ${\bf v}_s(t,{\bf x})$, e.g., via a decomposition
of it into elementary EPK processes, see eq.~(\ref{eq5_13}), becomes
essential.  Due to hydrodynamic confinement, the parameters $a_h$,
$b_k$, $\xi_h$, $\tau_{0,h}$ and $\lambda_k$ entering
eq.~(\ref{eq5_13}) do depend on the position ${\bf x}$. Substituting
the expansion eq.~(\ref{eq5_13}) into eq.~(\ref{eq6_2}) leads to
stochastic differential equations that are formally similar to
nonlinear Langevin equations \cite{nonlinlang}. With respect to the
Wiener-Langevin counterparts, they do not suffer all the troubles of
the Wiener singularity (lack of bounded variation) in the definition
of the stochastic integrals, forcing the choice of a stochastic
calculus (Ito, Stratonovich, H\"anggi-Klimontovich) by causing the
so-called Ito-Stratonovich dilemma \cite{vankampen}. Nevertheless,
their emergent properties could be far from trivial and could lead to
interesting new phenomena.  To clarify this issue, consider the
simplest but physically meaningful example of a micrometric particle
moving in a still liquid at constant temperature $T$ close to an
infinite wall, subjected to gravity and to a repulsive double layer
Debye potential from the wall, comprehensively described by the
potential $\phi({\bf x})$.  This problem has been analyzed
experimentally in \cite{volpe,volpe1}.  Indicate with $x$ the distance
of the particle from the wall so that $\phi=\phi(x)$ and,
  according to eqs.~(\ref{eq2_3}), (\ref{eq2_4a}),(\ref{eq5_13}),
\begin{eqnarray} 
\dot{x} & =  & v_d 
+\sum_{h=1}^{N_\Lambda}  a_h(x) \,  \Lambda_g(t,\xi_h(x),\tau_{0,h}(x);v)
+ \sum_{k=1}^{N_\Xi} b_k(x) \, \Xi_g(t,\lambda_k(x);v) \nonumber \\
\dot{v}_d  &=& L_d[v_d; x] + L_i[\dot{v}_d; x]
-\frac{1}{m} \partial_x \phi_x(x) \:.
\label{eq6_3}
\end{eqnarray}
Equation~(\ref{eq6_3}) represents the prototype of a stochastic
problem, i.e., the formulation of a detailed hydrodynamic
  description of fluid/particle interactions, here by using the
velocity splitting approach.  In the overdamped case, it is known that
the equilibrium probability distribution $p^*(x)$ is the classical
Boltzmannian density, $p^*(x)= A e^{-\phi(x)/k_B T}$. But is this
still true if the fluid inertial contributions expressed by the Basset
long-range force are accounted for?  Albeit it is likely to be the
case, a conclusive answer to this problem has not yet been given. A
violation of the Boltzmannian behaviour controlled by the potential
$\phi(x)$, i.e., a stationary density $p^*(x)=A e^{-\psi(x)/k_B T}$
with $\psi(x) \neq \phi(x)$, would imply the emergence of
fluctuational forces defined by the potential $\psi(x)-\phi(x)$
deriving from confinement and hydrodynamic effects, conceptually
analogous to the Casimir forces between metallic plates
\cite{casimir}.  This, and even more interesting problems (arising
from coupling effects whenever the tensorial structure of the
hydrodynamic resistance and inertial term is accounted for
\cite{brenner,procopio}) are posed by microfluidic applications to
fundamental statistical physics, once the hydrodynamic/thermal
fluctuations are included in detail.  The use of the velocity
splitting approach, leading to stochastic models of the form of
eq.~(\ref{eq6_2}), represents a feasible way to analyze them, at least
via direct stochastic simulations.


\end{document}